\documentclass[useAMS,usenatbib]{mn2e}
\usepackage{graphicx}

\title[]{Orientation and Rotational Parameters of  Asteroid 4179 Toutatis: New Insights
 from Chang'e-2's Close Flyby}

\author[Yuhui Zhao, Jianghui Ji \& Jiangchuan Huang et al.]
{Yuhui Zhao$^{1}$\thanks{zhaoyuhui@pmo.ac.cn},
Jianghui Ji$^{1}$\thanks{Email:jijh@pmo.ac.cn},
Jiangchuan Huang$^{2}$,
Shoucun Hu$^{1,3}$,
Xiyun Hou$^{4}$,   \newauthor
Yuan Li$^{5}$ and Wing-Huen Ip$^{5,6}$ \\
$^{1}$Key Laboratory of Planetary Sciences, Purple Mountain Observatory,
Chinese Academy of Sciences, Nanjing 210008,China \\
$^{2}$China Academy of Space Technology, Beijing 100094,
China\\
$^{3}$University of Chinese Academy of Sciences, Beijing 100049, China\\
$^{4}$School of Astronomy and Space Science, Nanjing
University, 210023, China\\
$^{5}$Space Science Institute, Macau
University of Science and Technology, Taipa, Macau\\
$^{6}$Institute of Astronomy, National Central University, Taoyuan, Taiwan}

\begin{document}

\date{Received * December *; in original form * October *}

\pagerange{\pageref{firstpage}--\pageref{lastpage}} \pubyear{2002}

\maketitle

\label{firstpage}

\begin{abstract}
In this work, we investigate the rotational dynamics of the
ginger-shaped near-Earth asteroid 4179 Toutatis, which was closely
observed by Chang'e-2 at a distance of $770\pm120~$ meters from the
asteroid's surface during the outbound flyby \citep{Huang2013} on 13
December 2012. A sequence of high-resolution images was acquired
during the flyby mission. In combination with ground-based radar
observations collected over the last two decades, we analyze these
flyby images and determine the orientation of the asteroid at the
flyby epoch. The 3-1-3 Euler angles of the conversion matrix from
the J2000 ecliptic coordinate system to the body-fixed frame are
evaluated to be $-20.1^\circ\pm1^\circ$, $27.6^\circ\pm1^\circ$ and
$42.2^\circ\pm1^\circ$, respectively. The least-squares method is
utilized to determine the rotational parameters and spin state of
Toutatis. The characteristics of the spin-state parameters and
angular momentum variations are extensively studied using numerical
simulations, which confirm those reported by \citet{Takahashi2013}.
The large amplitude of Toutatis' precession is assumed to be
responsible for its tumbling attitude as observed from Earth.
Toutatis' angular momentum orientation is determined to be described
by $\lambda_{H}=180.2^{+0.2^\circ}_{-0.3^\circ}$ and
$\beta_{H}=-54.75^{+0.15^\circ}_{-0.10^\circ}$, implying that it has
remained nearly unchanged for two decades. Furthermore, using
Fourier analysis to explore the change in the orientation of
Toutatis' axes, we reveal that the two rotational periods are 5.38
and 7.40 days, respectively, consistent with the results of the
former investigation. Hence, our investigation provides a clear
understanding of the state of the rotational dynamics of Toutatis.

\end{abstract}


\section{Introduction}
The Apollo-type near-Earth asteroid (4179) Toutatis was originally
discovered on 10 February 1934 and remained a lost asteroid until it
was once again detected by C. Pollas and colleagues on 4 January
1989 in Caussols, France. From a dynamical viewpoint, the asteroid
moves on an approximately 4:1 resonant orbit at a large eccentricity
with the Earth and has passed through a close encounter with the
Earth every four years since 1992 \citep{Whipple1993,Krivova1994}.
Dating back to the decadal years of near-Earth flybys of the
asteroid, the radar observations obtained from Arecibo and Goldstone
reveal that Toutatis appears to be an irregularly shaped asteroid
with two distinct lobes
\citep{Ostro1995,Hudson1998,Ostro1999,Ostro2002,Hudson2003}. Various
types of ground-based observations indicate that Toutatis is a
tumbling, non-principal axis (NPA)-rotating small body
\citep{Hudson1995,Ostro1999}. These effects, as observed in
Earth-approaching flybys, have also been reported based on optical
observations and extensive radar measurements \citep{Spencer1995,
Hudson1995,Ostro1999,Takahashi2013}.

The first near-Earth flyby for Toutatis occurred in December 1992 at
a distance of 0.242 AU, when the asteroid again came into view.
Optical observations were gathered from at least 25 sites around the
world through an international campaign. The observed rotational
light curves of Toutatis appeared to be highly unusual, with a large
amplitude and a non-periodic long rotation period. Subsequently,
\citet{Spencer1995} reported two major periods of complex rotation
of approximately 7.3 and 3.1 days, as estimated from analysis of the
data. In addition, they reported that Toutatis was the first
asteroid to show strong photometric evidence of complex rotation.
However, the authors did not clarify this complex rotation
phenomenon.

Furthermore, radar observations were performed by Goldstone in
California and by Arecibo Observatory during Toutatis' approach in
1992. The delay-Doppler images achieved a spatial resolution of 19
meters in range and 0.15 millimeters per second in radial velocity.
\citet{Ostro1995} suggested a rotational period between 4 and 5 days
based on these radar data. According to the investigations of
\citet{Burns1971,Burns1973}, the damping timescale for the slow
non-principal axis rotation of Toutatis exceeds the age of the solar
system. Thus, \citet{Ostro1995} noted that the spin state of
Toutatis may be primordial. However, recent investigation suggests
that YORP effects may slow the spin states of asteroids. Thus,
Toutatis' spin state remains a mystery.

Based on these high-resolution delay-Doppler radar observations,
\citet{Hudson1995} used a least-squares estimation to calculate
Toutatis' three-dimensional shape, spin states, and
moment-of-inertia ratios. They showed that the dimensions along the
three principal axes are 1.92, 2.40 and 4.60 kilometers and that
Toutatis rotates in a long-axis mode. The two major periods were
found to be 5.41 days for the rotation about the long principal axis
and 7.35 days for the long-axis precession about the angular
momentum vector. The results derived from the radar data were
inconsistent with the solutions presented by \citet{Spencer1995}.

Moreover, \citet{Hudson1998} adopted the published optical light
curves \citep{Spencer1995} and a radar-derived shape and spin-state
model \citep{Hudson1995} to estimate the Hapke parameters of
Toutatis. The Hapke photometric model was applied, and a $\chi_{2}$
minimization proposed by \citet{Hudson1997} was performed. The
synthetic light curves that were generated based on their model
provided a good fit to the optical data, with an rms residual of
0.12 mag. They showed that the combination of the optical data and
radar observations led to an estimation of the spin-state parameters
for Toutatis that was superior to the radar-derived outcomes. The
two parameters describing the moment-of-inertia ratios were
determined to be 3.22 and 3.02, respectively.

Based on the triaxial ellipsoid shape and spin state given by
\citet{Ostro1995}, \citet{Kryszczynska1999} presented the results of
modeling the light curve variations of this unusual rotating
asteroid by numerically integrating Euler's equation in combination
with the explicit expression for an asteroid's brightness as a
function of Euler angles. They achieved good agreement between the
observed and calculated light curves. They emphasized that the light
curves of Toutatis were dominated by the precession effect and by
the superposition of precession and rotation, which resulted in an
unapparent relationship between the rotation period alone and the
light curves. This understanding yielded an appropriate explanation
for the inconsistency between the rotational period of Toutatis
determined from optical data \citep{Spencer1995} and that determined
from radar observations \citep{Hudson1995}.

During the 1996 near-Earth approach, Toutatis was observed by the
Goldstone 8510-MHz radar system. Based on the physical model derived
from the observations of the 1992 approach, \cite{Ostro1999}
analyzed the radar measurements and refined the estimates of the
spin state of Toutatis. The combination of optical and radar data
was proven to better predict the orientational sequence displayed in
the images captured in 1996. After refinement, the two periods of
Toutatis were updated and estimated to be $5.376\pm0.001$ days for
the rotation about the long principal axis and $7.420\pm0.005$ days
for the uniform precession of the long principal axis about the
angular momentum vector. These two parameters yielded
moment-of-inertia ratios of $3.22\pm0.01$ and $3.09\pm0.01$. Thus,
the orientation at the 2004 approach could be predicted in both
inertial and geocentric coordinate systems.

\citet{Scheeres2000} determined that mutual gravitational
interactions between an asteroid and a planet or another asteroid
can play a significant role in shaping the asteroid's spin state.
They analyzed the interactions of a sphere with an arbitrary mass
and with Toutatis based on the radar-derived shape model. The
results thus obtained could partially explain the phenomenon of
Toutatis' current unusual rotational state. It was demonstrated that
the tumbling spin state of Toutatis might have been caused by
near-Earth flybys over its lifetime. This hypothesis enabled the
estimation of the mass distribution and moment-of-inertia for
Toutatis \citep{Busch2012}, thereby allowing the likely internal
structure to be inferred.

Using radar observations of five flybys from 1992 to 2008,
\citet{Takahashi2013} modeled the rotational dynamics and estimated
Toutatis' spin-state parameters using the least-squares method. They
calculated the Euler angles, angular velocities, and
moment-of-inertia ratios as well as the center-of-mass
(COM)-center-of-figure (COF) offset. By directly relating the
COM-COF offset and the moment-of-inertia ratios to the spherical
harmonic coefficients of the first- and second-degree gravity
potential, they could determine the driving force of the external
torque due to an external spherical body and evaluate the spin
state. The terrestrial and solar tidal torques were considered in
their dynamical models, and all aforementioned parameters were
included in the variable state vector to be estimated in the study.
Furthermore, the spin states and uncertainties were propagated to
the 2012 flyby epoch.

On 13 December 2012, the first space-borne close observation of
Toutatis was achieved by the second Chinese lunar probe, Chang'e-2,
at a distance of $770\pm120~(3\sigma)$ meters from Toutatis' surface
\citep{Huang2013}. Optical images of the asteroid were acquired by
one of the onboard engineering cameras during the outbound flyby.
Through analysis of over 400 images, \citet{Huang2013} estimated
Toutatis' osculating orbit, its dimensions along the major axes, and
its orientations. The highest resolution of the images was better
than 3 meters. New discoveries were made, including the presence of
a giant depression at the large end, a sharply perpendicular
silhouette near the neck region, and direct evidence of boulders and
regoliths. The geological features suggest that Toutatis may have a
rubble-pile structure. The physical length and width were determined
to be $4.75\times1.95~\rm{km}\pm~10\%$, respectively, and the
direction of the $+z$ axis was calculated to be (234.1$^\circ$,
60.7$^\circ$). They showed that the bifurcated configuration may
indicate that Toutatis is of contact binary origin and that it is
composed of two major lobes (head and body).

In this work, we perform an extensive investigation of the optical
images of Toutatis captured by Chang'e-2, and we determine the
orientation of the asteroid at the flyby epoch. In combination with
radar observations (\citet{Takahashi2013} and references therein),
we estimate the rotational parameters of Toutatis. Moreover, the
solar and terrestrial tidal torques are considered in the
establishment of the rotational dynamics model. The torque due to
the misalignment of the center of mass and the origin of the
body-fixed frame is evaluated to be insignificant at first order
\citep{Hudson2003,Busch2012,Busch2014}. Furthermore, we incorporate
the external gravitational tidal effects from the Moon and Jupiter
in our dynamical model.

Compared with the previous prediction \citep{Takahashi2013}, our
results for Toutatis' orientation, derived for Chang'e-2's flyby
epoch from both radar data and optical images, demonstrate good
consistency with the observational results of the spacecraft
\citep{Huang2013,Zou2014}. Our simulations reproduce the trajectory
of the long axis in space, with a precession amplitude of
approximately $60^{\circ}$. This high amplitude of Toutatis'
precession is supportive of its tumbling attitude as observed from
Earth. The characteristics of the angular momentum variations is
investigated in detail, and the variation induced by the near-Earth
flyby in 2004 is estimated to be $0.03\%$. The orientation of its
angular momentum in space is found to be described by
$\lambda_{H}=180.2^{+0.2^\circ}_{-0.3^\circ}$ and
$\beta_{H}=-54.75^{+0.15^\circ}_{-0.10^\circ}$, and therefore, this
orientation has remained nearly constant over the past two decades.
The rotational periods are estimated from the simulations to be 5.38
and 7.40 days for the rotation and precession, respectively. These
values are in good agreement with the work of \citet{Ostro1999}.

This work is structured as follows: Section 2 presents the
observational data, which comprise ground-based measurements and
optical images acquired by Chang'e-2. In this section, we also
analyze the optical data to derive the orientation of Toutatis at
the flyby epoch. In Section 3, we model the rotational dynamics of
Toutatis based on Euler's equation. The least-squares and multiple
shooting methods are employed to fit the variable state vector and
the corresponding results. The simulation results are presented in
Section 4. Finally, we conclude by discussing the innovations of our
investigation compared with previous works.

\section{Observations}

As described above, Toutatis progrades on an approximately 4:1
resonant eccentric trajectory with the Earth. Orbital determination
and rotational parameters for Toutatis have been documented since
the asteroid began to be continually observed in 1992. Since that
time, ground-based observations have been performed for its every
near-miss of Earth. As is well known, on 13 Dec 2012, Chang'e-2
completed the first successful close flyby of Toutatis and acquired
numerous images of this asteroid \citep{Huang2013}.

Using the released data from the Minor Planet Center and hundreds of
optical observations from the ground-based observational campaign
that lasted from July to December of 2012, the orbital determination
of Toutatis was precisely achieved within uncertainties on the order
of several kilometers, and the orbital parameters at the flyby epoch
were calculated to be $a$=2.5336 AU, $e$=0.6301, $i$=0.4466$^\circ$,
$\Omega$=124.3991$^\circ$, $\omega$=278.6910$^\circ$ and
$M$=6.7634$^\circ$. Hence, the initial orbit can be integrated to
calculate the relative positions of Toutatis with respect to the
Sun, Earth, Moon and other major planets in the solar system, which
are required for computing the external torques from the solar
tides, the terrestrial tides and the gravitational tides from other
bodies.

The positions of the major planets and the Moon are calculated based
on the DE405 ephemerides released by JPL \footnotemark[1]. The
gravitation of the Sun, the major planets and 67 asteroids in the
main belt as well as post-Newtonian effects are considered in the
dynamical model to achieve the orbital integration of Toutatis. In
addition, Chebyshev polynomial fitting is numerically implemented to
obtain the position of the asteroid at any given epoch.

\footnotetext[1]{{\citet{Takahashi2013} used the DE430
planetary ephemeris which is more accurate for fitting the
observation data of Toutatis. However, the position offset between
DE405 and DE430 is simply a few kilometers that will induce
systematic errors of approximately 10 parts per million into our torque
calculations, which is too small to change the conclusions of this work.}}

\subsection{Radar Measurements}
Radar measurements of Toutatis acquired by Goldstone and Arecibo
from 1992 to 2008 are used to solve for the asteroid's rotational
parameters. \citet{Takahashi2013} presented 33 sets of radar
observations. Together with the orientation obtained by Chang'e-2 at
the flyby epoch (see Section 2.2), we have 33 sets of ground-based
observation outcomes, including Euler angles, angular velocities and
one space-borne orientation parameter. The observational data are
summarized in Table 1. The observational errors of the radar data
are estimated to be between $3^{\circ}$ and $15^{\circ}$
\citep{Takahashi2013} for the Euler angles and between 2$^{\circ}$
day$^{-1}$ and 10$^{\circ}$ day$^{-1}$ for the components of angular
velocity; these errors are taken into account in our fitting.

\begin{table*} 
 \caption{Observational data for Toutatis from 1992 to 2012 (\citet{Takahashi2013} and references therein).}
\begin{tabular}{cccccccc}
\hline Year & Month & Date & Hour & Minute & Second & Euler angles
($^\circ$) & Angular velocity ($^\circ$/day) \\ \hline
1992 & 12 & 2 & 21 & 40 & 0 & (122.2, 86.5, 107.0) & (-35.6, 7.2, -97.0) \\
1992 & 12 & 2 & 19 & 30 & 0 & (86.3, 81.8, 24.5) & (-16.4, -29.1,-91.9) \\
1992 & 12 & 4 & 18 & 10 & 0 & (47.8, 60.7, 284) & (29.1, -23.2, -97.8) \\
1992 & 12 & 5 & 18 & 50 & 0 & (14.6, 39.4, 207.1) & (33.3, 8.2, -92.9) \\
1992 & 12 & 6 & 17 & 30 & 0 & (331.3, 23.7, 151.6) & (6.6, 34.5, -95.8) \\
1992 & 12 & 7 & 17 & 20 & 0 & (222.5, 25.4, 143.9) & (12.8, 25.4, -104.1) \\
1992 & 12 & 8 & 16 & 40 & 0 & (169.8, 45.5, 106.9) & (-31.1 -21.9, -97.7) \\
1992 & 12 & 9 & 17 & 50 & 0 & (137.3, 71.3, 22.3) & (11.8, -36.9, -94.9) \\
1992 & 12 & 10 & 17 & 20 & 0 & (103.1, 85.2, 292.6) & (35.8, -8.9, -97.9) \\
1992 & 12 & 11 & 9 & 40 & 0 & (77, 85.7, 225.5) & (31, 17, -96.3) \\
1992 & 12 & 12 & 9 & 20 & 0 & (42.8, 70.2, 133.2) & (-1.3, 37, -95.9) \\
1992 & 12 & 13 & 8 & 10 & 0 & (13.7, 44.4, 51.9) & (-38.3, 17.9, -97.3) \\
1992 & 12 & 14 & 7 & 50 & 0 & (323.7, 14, 0) & (-70.5, -50.6, -91.1) \\
1992 & 12 & 15 & 7 & 50 & 0 & (193.2, 24.4, 21.4) & (22.1, -26.6, -96.6) \\
1992 & 12 & 16 & 7 & 10 & 0 & (165.1, 46.4, 310.6) & (33.4, -3.4, -93.7) \\
1992 & 12 & 17 & 6 & 49 & 0 & (130.6, 76.1, 234.9) & (12.6, 33.9, -94) \\
1992 & 12 & 18 & 7 & 9 & 0 & (91.6, 81.6, 142.4) & (-24.3, 29.6, -102) \\
1996 & 11 & 25 & 19 & 48 & 0 & (130.5, 78.9, 143.2) & (-32, 16.4, -98.2) \\
1996 & 11 & 26 & 17 & 51 & 0 & (94.2, 88.1, 57.7) & (-30.6, -18.7, -91.5) \\
1996 & 11 & 27 & 17 & 34 & 0 & (60.4, 81.2, 320.9) & (10.7, -36.8, -94.7) \\
1996 & 11 & 29 & 15 & 37 & 0 & (349.3, 30, 168) & (23.1, 28.9, -98.3) \\
1996 & 11 & 30 & 15 & 47 & 0 & (250.3, 14.2, 166.9) & (-18.6, 32.1, -94.9) \\
1996 & 12 & 1 & 14 & 23 & 0 & (180.4, 37.6, 139.3) & (-38.7, -0.5, -98.1) \\
1996 & 12 & 2 & 13 & 43 & 0 & (146.7, 64, 64.9) & (-12.6, -34.8, -97.9) \\
1996 & 12 & 3 & 12 & 20 & 0 & (116.7, 81.4, 340.4) & (24.3, -28.2, -98.1) \\
2000 & 11 & 4 & 17 & 6 & 0 & (110, 88.5, 30) & (0, -32.5, -98.9) \\
2000 & 11 & 5 & 18 & 1 & 0 & (70.6, 84, 281) & (34.5, -17.2, -97.9) \\
2004 & 10 & 7 & 13 & 56 & 0 & (79.19, 85.3, 365.2) & (-2.5, -35.4, -109) \\
2004 & 10 & 8 & 14 & 4 & 0 & (44.9, 72.5, 263.1) & (32.4, -18.1, -97.9) \\
2004 & 10 & 9 & 13 & 57 & 0 & (12.8, 47.3, 181.4) & (29.7, 22.8, -98.1) \\
2004 & 10 & 10 & 13 & 17 & 0 & (327.7, 20.4, 124.1) & (-10.7, 34.7, -97.3) \\
2008 & 11 & 22 & 10 & 54 & 0 & (119.5, 90.7, 92) & (118.1, 90.4, 93.6) \\
2008 & 11 & 23 & 10 & 45 & 0 & (86.2, 85, 0.3) & (-0.4, -36.2, -98.9) \\
2012 & 12 & 13 & 8 & 30 & 0 & (-20.1, 27.6, 42.2) &  \\ \hline
\end{tabular}
\end{table*}

\subsection{Observations by Chang'e-2}
As mentioned previously, Chang'e-2 captured Toutatis' silhouette via
one of the onboard engineering cameras at the time when the asteroid
was approaching the Earth in December 2012. The camera has a lens
with a 54-mm focal length and a 1024-by-1024-pixel CMOS detector.
The field of view of the camera is $7.2^\circ$ by $7.2^\circ$. The
images were acquired during the outbound flyby of Chang'e-2 because
of the large Sun-Toutatis-Chang'e-2 phase angle on the inbound route
\citep{Huang2013,Zhao2014a}. The imaging of Toutatis lasted
approximately 25 minutes. \citet{Huang2013} reported the first
panoramic image of the asteroid in the sequence, which was acquired
at a distance of 67.7 km from Toutatis at a resolution of 8.30 m, as
shown in Figure 1a.
\begin{figure*}
\includegraphics[scale=0.70]{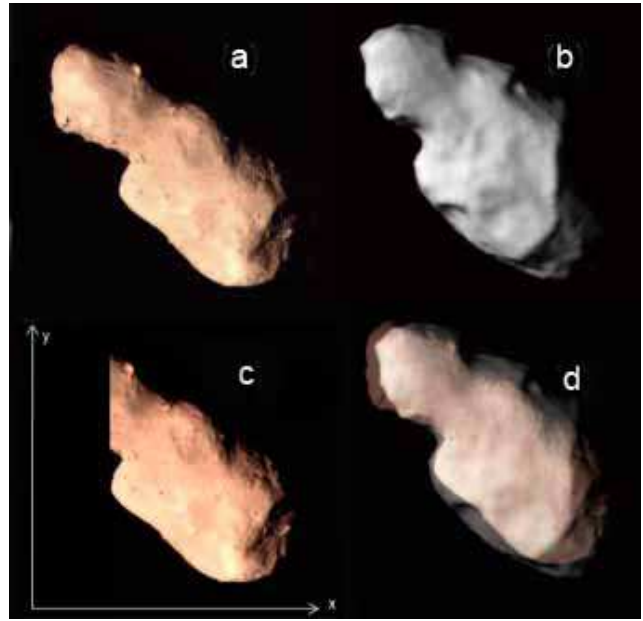}
\centering \caption{Images of Toutatis. a: First panoramic image
captured by Chang'e-2. b: Best-matching attitude of the radar model
with the optical image shown in Fig. 1a. c: Illustration of the
graphical frame of the camera. d: Combination and comparison of the
radar model with the optical results. }
\end{figure*}

\subsubsection{Attitude matching}
The 3D shape model derived from delay-Doppler radar imaging, as
presented in Figure 1b, is used to discern the attitude of Toutatis.
The latest radar-derived shape model \citep{Busch2012} was
constructed based on additional radar measurements performed by
Goldstone in 2000 and by Arecibo in 2004 and 2008, and in this work,
this model is employed in combination with the optical images from
Chang'e-2's flyby to match the spin state of Toutatis.

In general, the attitude of a rigid body in space can be determined
from its rotations about the three axes of an orthogonal coordinate
system. As Figure 2 shows, to obtain the attitude of Toutatis, we
suppose that the axes $l_{1}$, $l_{2}$ and $l_{3}$ are defined as
follows: the mutually perpendicular axes $l_{1}$ and $l_{2}$ extend
through the center of Toutatis' shape and along the directions of
the long short axes in the image, respectively. In addition, $l_{3}$
is perpendicular to the image plane through the intersection of
$l_{1}$ and $l_{2}$, and thus, these axes form a right-handed
coordinate system.

Considering the render and the orientation of the camera's optical
axis \citep{Zhao2014a}, the 3D radar-derived shape model of Toutatis
can be rotated at an interval of $1^{\circ}$ for each of the three
Euler angles about the three principal axes of its body-fixed frame
to match its attitude to that shown in the optical images. As shown
in Figure 2, we choose three criteria related to the optical image
to determine whether the rotated model is consistent with the
optical results from Chang'e-2's view direction: (1) the slope of
the long axis, represented by the red line in Figure 2a; (2) the
ratio of the long axis to the short axis, indicated by the green
line; and (3) the obvious topography on the neck area connecting the
two major lobes of the asteroid, as shown in Figure 2b and 2c. These
features of the optical image can be reproduced by rotating the 3D
radar shape model about its three principle axes. Figure 1b shows
the best approximation of the attitude of the radar model to that
indicated by the optical image shown in Figure 1a. The two images
are quantitatively compared in Table 2.

\begin{table}
 \begin{minipage}{90mm}
  \caption{Comparison of optical and radar image results.}
\begin{tabular}{lll}

\hline Property            & Optical Image & Radar Image \\ \hline
Slope                      &  1.386  &1.340 \\
Ratio of length to width   &  2.52   &2.08 \\
\hline \label{comp}

\end{tabular}
\end{minipage}
\end{table}

Between the optical image and the model, the difference in the slope
of $l_{1}$ is not obvious; however, the deviation in the ratio of
the length to the width appears to be significant. \citet{Zou2014}
suggested using a render frame with the lighting of the radar model
and combining multiple optical images using computer graphics
methods, which may yield a likely explanation for the higher value
of the length-to-width ratio obtained for the optical images
acquired by Chang'e-2.

Using the optical images from Chang'e-2, \citet{Bu2015} rotated the
radar-derived model and retrieved an orientation with respect to the
graphical frame described by direction cosine angles of
$(126.13\pm0.29^{\circ}, 122.98\pm0.21^{\circ},
126.63\pm0.46^{\circ})$. In accordance with the definition of the
cosine angles given by \citet{Bu2015}, we calculate this set of
cosine angles to be $(130.3\pm1.0^{\circ}, 134.78\pm1.0^{\circ},
106.95\pm1.0^{\circ})$. The two sets of results are quite similar to
each other. However, we should note that the results presented here
neglect the attitude of the camera in the inertial frame.

\begin{figure*} 

\includegraphics[scale=0.33]{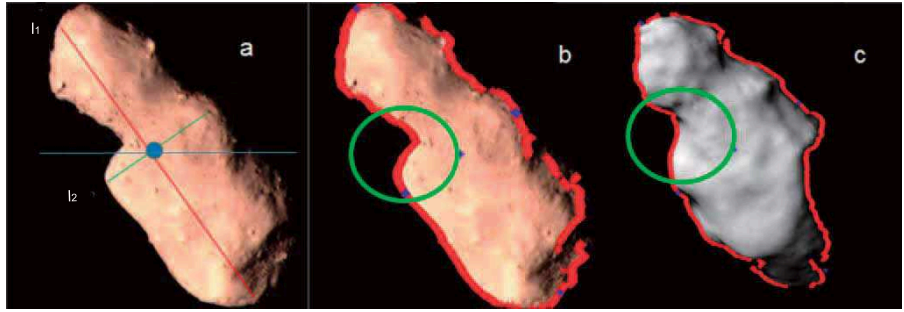}
\centering \caption{Properties of an optical image of Toutatis. a:
The red line and green line represent the lengths of the image of
the asteroid along the $l_{1}$ and $l_{2}$ directions, respectively;
the ratio between these lengths is one of the properties that is
used to characterize the image. The blue line represents the
horizontal line, which can be used to determine the slope of
$l_{1}$. b and c: The area within the green circle represents an
obvious characteristic topological feature that can be used to
derive information concerning the rotation about $l_{1}$. }
\end{figure*}

\subsubsection{Euler angles}
Because of the fixed direction of the camera's optical axis,
Chang'e-2 maintained a nearly constant attitude throughout the
shooting process. Figure 3 depicts the spacecraft's body-fixed
frame, where $\vec{l}$ is the direction of the camera's optical axis
and the corresponding unit vector in the spacecraft's body-fixed
frame is $(-0.06976, 0.9976, 0)$. The location relationship
represents a transformation from Chang'e-2's body-fixed frame to the
graphical frame shown in Figure 1c. In combination with the attitude
information of the spacecraft, we can establish a relationship
between the graphical frame and the inertial coordinate system. As
Figure 1c shows, the left portion of Toutatis is blocked by the
solar panel on the spacecraft. The unit vectors of the x and y axes,
which represent the horizontal and vertical axes in the graphical
frame, are determined to be $(0.824,0.540,-0.173)$ and
$(0.344,-0.233,0.910)$, respectively, in the J2000 equatorial
coordinate system \citep{Huang2013b}.

\begin{figure*}
\includegraphics[scale=0.50]{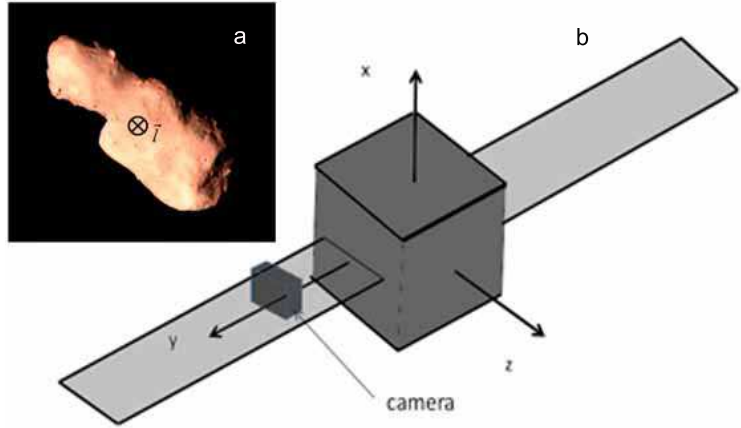}
\centering \caption{a. Graphical frame. b. Spacecraft's body-fixed
frame.}
\end{figure*}

By merging the rotated radar-derived shape model and the optical
image, Figure 1d illustrates the similarities and differences
between the two models. In combination with the conversion
relationship among the asteroid's body-fixed frame, the graphical
frame, and the inertial coordinate system, we can obtain the matrix
describing the transformation from the J2000 ecliptic coordinate
system to the asteroid's body-fixed system expressed in terms of
3-1-3 Euler angles as below:
\begin{equation} 
\vec{R}=R_z(42.2^\circ)R_x(27.6^\circ)R_z(-20.1^\circ)\vec{r},
\end{equation}
where $R_x$ and $R_z$ are the standard rotation matrices for
right-handed rotations around the $X$ and $Z$ axes, respectively.
The coordinate transformation and the corresponding Euler angles
will be briefly introduced in the next section. The orientation of
the principle axis is then obtained with respect to the attitude of
the radar model, which is estimated to be ($249.87 \pm 1^\circ,
62.43 \pm1^\circ$) in the J2000 ecliptic coordinate system
\citep{Zhao2014b}. Corresponding errors arise from the matching
process, the uncertainties of the radar-derived shape model and the
attitude uncertainties of the spacecraft \citep{Huang2013}.

\section{Numerical Models}
\subsection{Dynamical Model}
The rotation matrix shown in equation ($1$), which is composed of
the three 3-1-3 Euler angles $\vec\alpha=(\alpha,\beta,\gamma)$,
maps the information for conversion from the inertial coordinate
system to the body-fixed frame, as shown in Figure 4. The body-fixed
frame is generated by applying the following rotation sequence of
the yaw, pitch and roll angles: a rotation of $\alpha$ around the z
axis, then a rotation of $\beta$ around the x axis, and finally, a
rotation of $\gamma$ around the z axis from the inertial frame.

\begin{figure*} 
\includegraphics[scale=0.48]{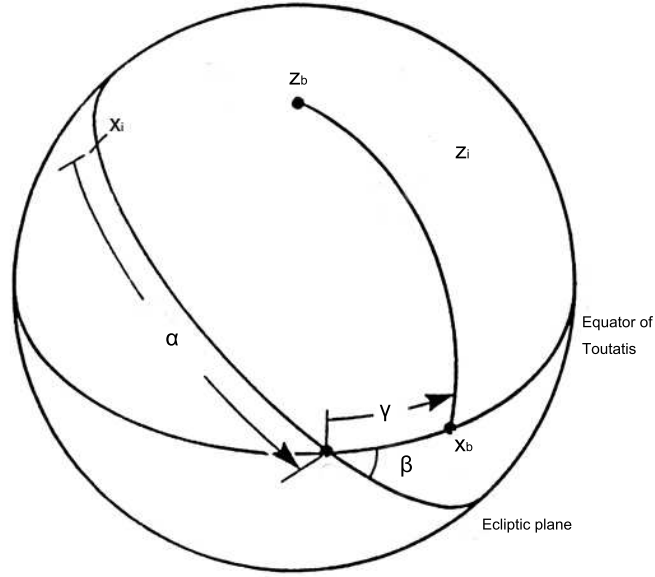}
\centering \caption{Coordinate system transformation relationship in
terms of 3-1-3 Euler angles. Subscripts i and b indicate the
inertial coordinate system and the body-fixed frame, respectively.
The three angles $\alpha, \beta,$ and $\gamma $ form a 3-1-3 set of
Euler angles. }
\end{figure*}

The Euler angles describe the orientation of a rigid body in space
at a specific time, and their variations represent spin states. Let
the vector $\vec{\omega}$ define the instantaneous rotational
velocity in the body-fixed frame. Then, the set of kinematic
differential equations for the 3-1-3 Euler angles is as follows:
\begin{eqnarray} 
\dot{\vec{\alpha}}=\frac{1}{\sin\beta} \left(\begin {array}{c}
~~~~\sin\gamma~~~~~~~~~~~\cos\gamma~~~~~~0 \\
\cos\gamma\sin\beta~~~-\sin\gamma\sin\beta~~0\\
-\sin\gamma\cos\beta~~-\cos\gamma\cos\beta ~~\sin\beta
\end{array}\right)\vec{\omega} \nonumber \\
=[B(\vec{\alpha})]\vec{\omega} ~~.
\end{eqnarray}

The time derivative of the Euler angles encounters a singularity at
either $\beta=0^{\circ}$ or $\beta=180^{\circ}$, which may cause
$\alpha$ and $\gamma$ to rotate in the same plane.

As derived from the Euler equation, Euler's rotational equation of
motion describes the time derivative of the angular velocities:

\begin{equation} 
[I]\dot{\vec{\omega}}=-[\tilde{\omega}][I]\vec{\omega}+\vec{L} ~~,
\end{equation}
where the moment-of-inertia matrix $[I]$ is constant, symmetric and
given in the body-fixed frame. It has dimensions of $[3\times3]$ and
can be defined in terms of six quantities. $\vec{L}$ represents the
external torques acting on the dynamical system, and the matrix
$[\tilde\omega]$ has the following form:

\begin{equation} 
[\tilde\omega]= \left(\begin {array}{c}
~~0~~~~~~~-\omega_{3}~~~\omega_{2} \\
~~\omega_{3}~~~~~~~0~~~~-\omega_{1}\\
-\omega_{2}~~~~~\omega_{1}~~~~0
\end{array}\right) ~~.
\end{equation}

Then, the time derivative of the angular velocities is expressed as
follows:

\begin{equation} 
\dot{\vec{\omega}}=[I]^{-1}(-[\tilde{\omega}][I]\vec{\omega}+\vec{L})
~~.
\end{equation}

To calculate the external torque exerted by a spherical body, we
assume that $\vec{r}'$ is the vector of an infinitesimal mass
element $(dm)$ of the asteroid relative to the origin. Figure 5
presents a schematic diagram of the effect exerted by a spherical
perturbing body on an irregularly shaped rigid body. According to
the definition of angular momentum $\vec{H}$, we have

\begin{equation} 
\frac{\rm{d}\vec{H}}{\rm{d}t}=\frac{\rm{d}}{\rm{d}t}\int dm~\vec{r}'\times\dot{\vec{r}'}
=\int
\vec{r}'\times(d\vec{F}_{G}/dm-\ddot{\vec{B}})dm ~~,
\end{equation}
where $\ddot{\vec{B}}$ is the acceleration of the origin in the
inertial coordinate system and $d\vec{F}_{G}$ is the gravitational
attraction experienced by an arbitrary mass element. When the center
of mass of the rigid body is chosen as the origin $\bar{O}$, we have
$\int dm \vec{r}'\times\ddot{\vec{B}}=0$. Therefore, the torques
exerted on the rigid body are written in the form \citep{Schaub2009}

\begin{equation} 
\vec{L}_{c}=\int_{M}\vec{r}'\times d\vec{F}_{G} ~~.
\end{equation}
Otherwise, if the vector of $(dm)$ relative to the center of
mass is $\vec{r}_{c}$, equation (6) has the following form:

\begin{eqnarray} 
\frac{\rm{d}\vec{H}}{\rm{d}t}
=\int\vec{r}'\times(d\vec{F}_{G}/dm-\ddot{\vec{B}})dm \nonumber \\
=\int(\vec{r}_{c}-\vec{r}_{cm})\times(d\vec{F}_{G}/dm-\ddot{\vec{B}})dm
~~,
\end{eqnarray}
where $\vec{r}_{cm}=(x_{cm},y_{cm},z_{cm})^T$ is the vector of the
center of figure with respect to the center of mass (the COM-COF
offset). Thus, equation (6) can be expressed as follows:
\begin{equation} 
\vec{L}=\int_{M}\vec{r}'\times
d\vec{F}_{G}+M\vec{r}_{cm}\times\ddot{\vec{B}} ~~,
\end{equation}
where $M$ is the total mass of the asteroid.

Let $\vec{r}$ be the position vector of the spherical body relative
to the asteroid. $\vec{R}=\vec{r}'-\vec{r}$ denotes the position of
a body element measured from the spherical body. Consequently,
$d\vec{F}_{G}$ has the following form:

\begin{equation} 
d\vec{F}_{G}=-\frac{GM_{s}}{R^{3}}\vec{R}dm ~~,
\end{equation}
where $M_s$ is the mass of the external spherical body. Hence, the
integral in equation (9) can be expressed as

\begin{equation} 
\vec{L}=-GM_{s} \int_{M}\vec{r}'\times \frac{\vec{R}}{R^{3}}dm +M\vec{r}_{cm}\times\ddot{\vec{B}}~~.
\end{equation}

\begin{figure*} 
\includegraphics[scale=0.40]{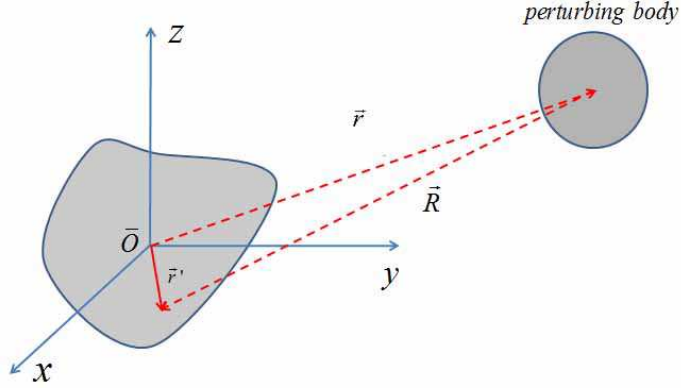}
\centering
\caption{External torque of a spherical body on an irregularly
shaped rigid body. }
\end{figure*}

As Figure 5 shows, under the assumption that $\vec{r}\gg\vec{r}'$, the first-order expansion of $\frac{\vec{R}}{R^{3}}$ has the following form:

\begin{equation} 
\frac{\vec{R}}{R^{3}}=-[1+\frac{3(\vec{r}\cdot\vec{r}')}{r^{2}}]
\frac{\vec{r}}{r^{3}}+\frac{\vec{r}'}{r^{3}} ~~.
\end{equation}

Substituting (12) into equation (9) allows the moment from the perturbing body
to be divided into two components:

\begin{equation} 
\vec{L}=\vec{L}_{1}+\vec{L}_{2} ~~,
\end{equation}
where
\begin{equation} 
\vec{L}_{1}=GM_{s}\int_{M}\vec{r}'\times\frac{\vec{r}}{r^{3}}dm+M\vec{r}_{cm}\times\ddot{\vec{B}} ~~,
\end{equation}
\begin{equation} 
\vec{L}_{2}=\frac{3GM_{s}}{r^{5}}\int_{M}
(\vec{r}\cdot\vec{r}')(\vec{r}'\times\vec{r})dm ~~.
\end{equation}
If the origin $\bar{O}$ is the center of mass of the irregularly
shaped rigid body, then we obviously have $\vec{r}_{cm}\equiv0$ and
$\vec{L}_{1}\equiv0$. Otherwise, truncated at the lowest order, we
have
\begin{eqnarray} 
\ddot{\vec{B}}=\frac{GM_{s}}{M}\int_{M}\frac{\vec{r}}{r^{3}}dm ~~.
\end{eqnarray}

If perturbing bodies other than the Sun are considered, then equation
(16) should incorporate their contributions. Substituting equation
(16) into equation (14), we obtain

\begin{eqnarray} 
\vec{L}_{1}=GM_{s}\int_{M}(\vec{r}'+\vec{r}_{cm})
\times\frac{\vec{r}}{r^{3}}dm \nonumber \\
=GM_{s}\int_{M}\vec{r}_{c}\times\frac{\vec{r}}{r^{3}}dm=0 ~~.
\end{eqnarray}

Essentially, the torques imposed by the COM-COF offset are far too
small to contribute in the first-order approximation. Furthermore,
the change in angular momentum that results from $L_{1}$ is not an
intrinsic feature of the rigid body; rather, it is determined solely
by the selection of the base point when solving for the rotational
dynamics.

\citet{Takahashi2013} expressed the second-degree potential as
follows:

\begin{equation} 
U_{2}=\frac{G}{2r^{3}}I_{T}-\frac{3G}{2r^{5}}\vec{r}[I]\vec{r} ~~,
\end{equation}
where $I_{T}$ is the trace of $[I]$. The partial derivative of the
potential can be derived as follows:
\begin{equation} 
\frac{\partial{U_{2}}}{\partial\vec{r}}
=-\frac{3G}{2r^{5}}I_{T}\vec{r}+\frac{15G}{2r^{7}}(\vec{r}[I]\vec{r})\vec{r}
-\frac{3G}{r^{5}}[I]\vec{r} ~~;
\end{equation}
therefore, the deduced moment can be expressed in the following
form:

\begin{equation} 
L_{2}=\frac{3GM_{s}}{r^{5}}[\tilde{r}][I]\vec{r} ~~.
\end{equation}
Our second-degree expansion of the moment given in equation (15) is
consistent with the results represented by equation (20), which
indicates that tidal torques caused by the oblateness of the rigid
body.

\subsection{Numerical Method}
The least-squares and multiple shooting methods are used to fit the
observational data and to simulate the propagation of the rotational
parameters. The dynamical equation of the state vector, which is
composed of three Euler angles and three angular velocities, is
written as follows:

\begin{equation} 
\dot{X}=F(X,t) ~~,
\end{equation}
where $ X=(\alpha,\beta,\gamma,\omega_{x},\omega_{y},\omega_{z})^T
$; $\alpha$, $\beta$, and $\gamma$ are the 3-1-3 Euler angles, and
$\omega_{1}$, $\omega_{2}$, and $\omega_{3}$ are the three
components of angular velocity. The previous results for $I_{xx}$,
$I_{yy}$, $I_{zz}$, $I_{xy}$, $I_{xz}$, and $I_{yz}$ are used in our
calculations \citep{Takahashi2013}. $F$ is the function that
represents the time derivative of $X$ and the dynamical model. In
the first-order approximation, the dynamical model has the following
form:

\begin{equation} 
\dot{X}=A(X,t)X(t) ~~,
\end{equation}
where $A(X,t)$ is the dynamical matrix. The transition matrix $\Phi$
is defined as follows:

\begin{equation} 
\Phi(t_{1},t_{2})=\frac{\partial x(t_{1})}{\partial x(t_{2})} ~~.
\end{equation}

Based on the nominal orbit, we establish the relationship between
the state vectors at two specific times, $t_{1}$ and $t_{2}$, as
follows:

\begin{equation} 
X(t_{2})=\phi(t_{1},t_{2})X(t_{1}) ~~.
\end{equation}

Then, the transition matrix can be obtained by integrating the
derivate equation:

\begin{equation} 
\dot\phi(t_{0},t)=A(t,X)\phi(t_{0},t) ~~.
\end{equation}

The observational equation for the measurements $Y$ is

\begin{equation} 
Y_{i}=Z(X_{i},t_{i})+\epsilon_{i} ~~,
\end{equation}
where $Z$ represents the observational model, the subscript $i$
indicates the sequence of the observational data and $\epsilon$
represents the observational uncertainties. In the first-order
approximation, the partial derivatives of the observational
quantities with respect to the variables form the relationships
between them, and we have

\begin{equation} 
Y_{i}=\frac{\partial{Z}}{\partial{X}}|_{x=x_{i}}X_{i}+\epsilon_{i}
=\frac{\partial{Z}}{\partial{X}}|_{x=x_{i}}\phi(t_{0},t_{i})X(t_{0})+\epsilon_{i}~~,
\end{equation}
where $t_{i}$ is the observation epoch of the data set. The
least-squares method adopted herein is similar to that of
\citet{Takahashi2013}. The cost function is defined as follows:
\begin{eqnarray} 
J= \frac{1}{2}(Y-\frac{\partial{Z}}{\partial{X}}|_{x=x_{i}}
\phi(t_{0},t_{i})X(t_{0}))^{T}W \nonumber \\
(Y-\frac{\partial{Z}}{\partial{X}}|_{x=x_{i}}
\phi(t_{0},t_{i})X(t_{0}))\nonumber \\
+\frac{1}{2}(\bar{X}(t_{0})-X(t_{0}))^{T}\bar{P}^{-1} \nonumber \\
(\bar{X}(t_{0})-X(t_{0}))\nonumber ~~,\\
\end{eqnarray}
where $W$ is the weighting matrix, and $P$ is the covariance matrix.
The bars over $X$ and $P$ represent the a priori values deduced
through estimation or from previous results. The modified
differential equation that computes the correction to the variables
can be expressed as follows:
\begin{eqnarray} 
\triangle X(t_{0})=\nonumber \\
(\sum H^{T}WH+\bar{P}^{-1})^{-1}(\sum
H^{T}WY+\bar{P}^{-1}\bar{X}(t_{0}))^{-1} ~~.
\end{eqnarray}

The RKF78 integrator can be used to integrate the rotational
equations. The initial step size is set to approximately $10^{-9
~\circ}$ for the Euler angles and $10^{-8~\circ}/day$ for the
rotational velocities. In the integration, an adaptive step size is
used to numerically solve the equations. The truncation errors for
the Euler angles and rotational velocities are set to
$10^{-8~\circ}$ and $10^{-7~\circ}/day$, respectively.

\section{Results}
To better understand the rotational dynamics of Toutatis, we
performed a large number of numerical simulations based on
Chang'e-2's observations and ground-based radar measurements, using
our dynamical models described above. In the following, we will
present the major outcomes for the spin states, rotational period
and variation in angular momentum of Toutatis.

In this work, the initial variables for Toutatis' spin state that
were used in our numerical simulations were adopted from
\citet{Takahashi2013} for the epoch $t_{0}$ (17:49:47 UTC on 9 Nov
1992). The orientation of the angular momentum and the rotational
periods were also calculated in the study. In our calculation, we
balanced the weights of the optical data based on the uncertainties
of the observations. Consequently, we derived innovative solutions
for the spin state of Toutatis, which are summarized in Table 3.

\begin{table*}
\caption{Spin-state parameters of Toutatis derived from our
numerical simulations.}
\begin{tabular}{ll}
\hline Property & Value\\ \hline
Simulated solutions at $t_{0}$ & $\alpha=147.5^\circ,\beta=63.9^\circ,\gamma=241.5^\circ$ \\
&$\omega_{1}=14.5^\circ/day,\omega_{2}=33.7^\circ/day, \omega_{3}=-98.5^\circ/day$\\
Results at flyby epoch & $\alpha=-3.65^\circ,\beta=43.62^\circ,\gamma=24.7^\circ$ \\
Orientation of angular momentum &
$\lambda_{H}=180.2^{+0.2^\circ}_{-0.3^\circ},
\beta_{H}=-54.75^{+0.15^\circ}_{-0.10^\circ}$\\
Rotational/precession period & $5.38$ days, $7.40$ days\\
\hline
\end{tabular}
\end{table*}

The residuals of the Euler angles (34 sets) and angular velocity (33
sets) were normalized with respect to the maximum radar
observational errors ($15^{\circ}$ for the Euler angles and
$10^{\circ}$/day for the angular velocities) and are shown in Figure
6. Because the previous prediction of the orientation at the
Chang'e-2 flyby epoch based on the radar-derived results differs
from that observed by Chang'e-2, the use of the optical data might
have degraded the convergence of the simulation algorithm.
Therefore, the magnitude of the residuals is larger than that
observed in the results from the radar data (Takahashi et al, 2013).
The residual errors in the simulations were normalized with respect
to the observational uncertainties. Because of the inconsistency of
the observational data, the magnitudes of the residuals are slightly
higher than those of the previous results. However, all deviations
lie within the $3\sigma$ region. The largest bias is found in the
roll angle, which exhibits a remarkable difference between the
prediction obtained from the radar measurements and the authentic
spin state of Toutatis that is directly indicated by Chang'e-2's
observations at the flyby epoch.

\begin{figure*} 
\includegraphics[scale=0.65]{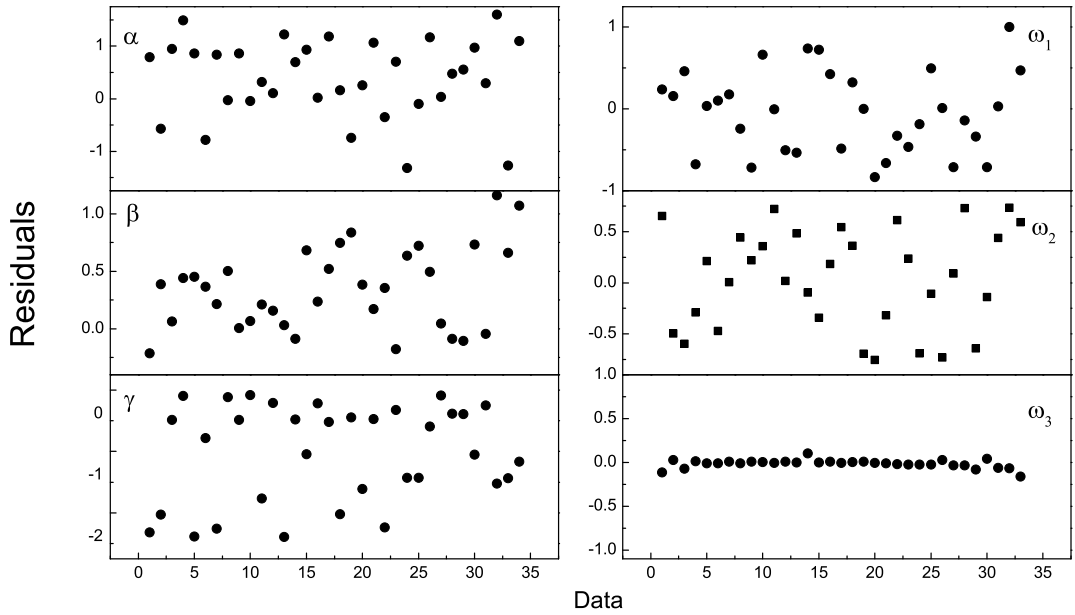}
\centering \caption{Residuals of rotational parameters with respect
to the maximum radar observational errors ( $15^{\circ}$ for the
Euler angles and $10^{\circ}$/day for the angular velocities). Left
panels: residuals of the Euler angles. Right panels: residuals of
the angular velocity components.}
\end{figure*}

\subsection{Spin States}
According to the numerical results derived from radar observations
collected before 2008 \citep{Busch2012}, we considered a render
effect for Toutatis and generated a predicted imaging outcome prior
to Chang'e-2's flyby, as shown in Figure 7a
\citep{Busch2012,Zhao2014a}. In addition, based on the images
acquired by Chang'e-2, we corrected the attitude of Toutatis by
rotating the radar-derived shape model (see Section 2.2.1) to search
for a good match with the Chang'e-2 images acquired at the flyby
epoch (Fig. 7b), which provide the only space-borne optical data
regarding Toutatis' orientation. Furthermore, the present
simulations yielded another solution for Toutatis' attitude during
the near-Earth flyby in 2012. Figure 7c shows the outcomes derived
from our rotational model using space- and ground-based
observations.

In comparison with the results obtained from the optical images
(Fig. 7b), the radar-derived results (Fig. 7a) exhibit a dramatic
deviation in the roll angle; hence, these results yield a different
profile of the asteroid. The simulation results derived from our
dynamical model (Fig. 7c) differ from those of Figure 7b with a
pitch angle bias of within $20^\circ$. Thus, we may safely conclude
that our outcomes represent a good improvement in the understanding
of Toutatis' spin state.

\begin{figure*} 
\includegraphics[scale=0.35]{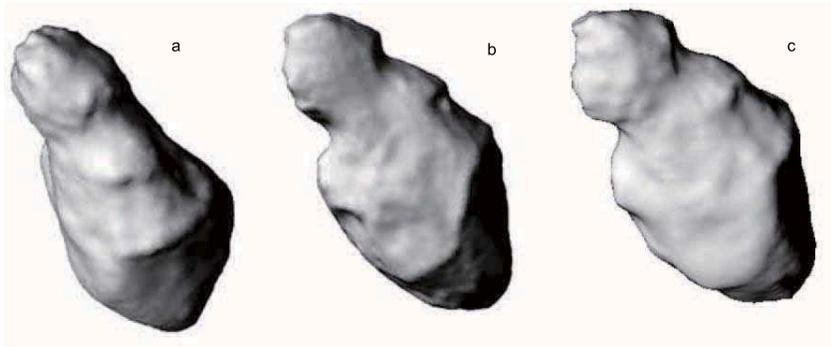}
\centering \caption{Comparison of Toutatis' orientation. a: The left
panel shows the results derived from \citet{Takahashi2013}'s work,
with an uncertainty of dozens of degrees. b: The middle panel shows
the results calculated from optical images acquired by Chang'e-2 and
by rotating the radar shape model. c: The right panel presents the
outcomes derived from our rotational model using space- and
ground-based observations, which are close to those obtained from
the optical images acquired by Chang'e-2.}
\end{figure*}

The orientation of the long axis in the inertial frame likely
reflects the precession of Toutatis. Based on the dynamical model of
rotation, we calculated the variation in the direction of the long
axis. Figure 8 shows the trajectories of the long axis with respect
to the J2000 ecliptic coordinate system in a unit sphere over the
past two decades. The motions of the long axis are projected onto
the X-Y, X-Z and Y-Z planes (see Figs. 8a, 8b and 8c, respectively).
The figure reveals that the long-axis motion of the asteroid has
remained ellipsoidal in the X-Y and Y-Z planes, whereas it has
rectilinearly precessed in the X-Z plane. All curves lie outside the
ecliptic plane, implying that the small lobe of Toutatis is always
located above the large end from a viewpoint close to the ecliptic.

Moreover, the orientation of the center axis of precession of
Toutatis can be approximately determined from Figure 8, and the
derived spherical coordinates can be estimated to be
$(-0.2^\circ,54.6^\circ)$ in the ecliptic coordinate system. As a
result of Toutatis' clockwise rotation and precession, the center
axis of precession points nearly along the opposite direction to the
angular momentum (see Section 4.3). The amplitude of the precession
is approximately $60^{\circ}$, which may shed light on the
significantly different attitudes of the asteroid that have been
observed from Earth.

\begin{figure*} 
\includegraphics[scale=0.40]{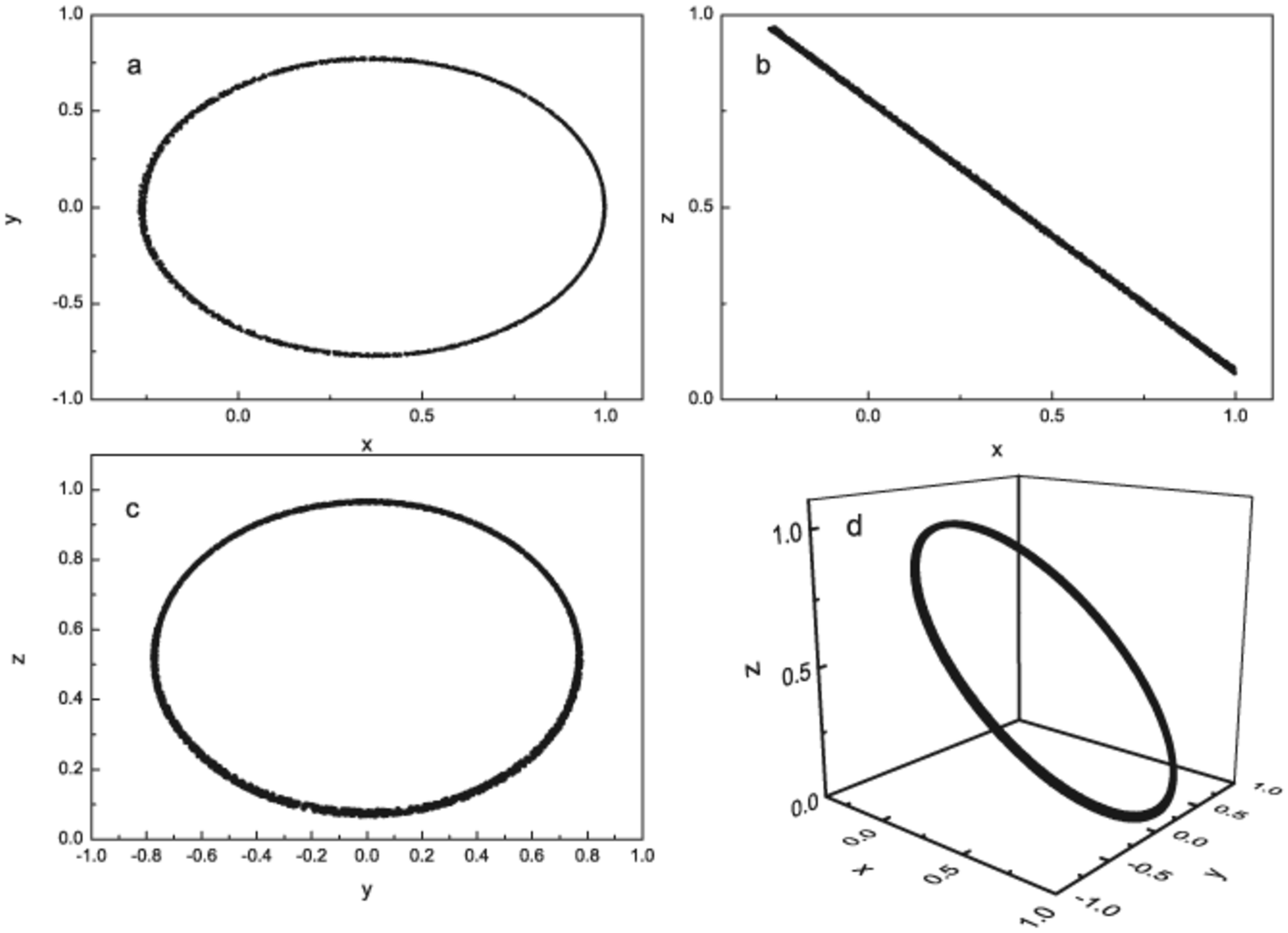}
\centering \caption{Trajectories in the J2000 ecliptic coordinate
system of the long axis of Toutatis in a unit sphere. Panels a, b,
and c show the trajectories in the X-Y, X-Z, and Y-Z planes,
respectively, and panel d shows the motion of the long axis in 3D.}
\end{figure*}

\subsection{Angular Momentum}
Considering the radar-derived shape model and the components of the
inertial matrix inferred by \citet{Takahashi2013}, we will now
explore the angular momentum of Toutatis induced by various external
gravitational torques. Figure 9 shows the variations in the external
gravitational torques acting on the spin state of Toutatis from 1992
to 2012. The solar torque is on the order of $10^{-10}-10^{-7}$,
indicating that its value is 2-3 orders higher at the perigee than
at the apogee. Its periodic variation is clearly associated with
Toutatis' orbital period. The variation tendencies of the
gravitational torques arising from the Earth and Moon are similar,
as shown by the red and blue curves, respectively. There is a
$10^{-2}$ difference in the orders of these torques because of the
magnitudes of the masses of the bodies from which they arise. The
periods of both torques are consistent with that of the black curve
because of Toutatis' resonance orbit with the Earth. At present,
Toutatis is also in a 3:1 mean motion resonance orbit with Jupiter
\citep{Whipple1993}; thus, one entire period of the torque induced
by Jupiter is displayed by the green curve \citep{Busch2014}.

\begin{figure*} 
\includegraphics[scale=0.55]{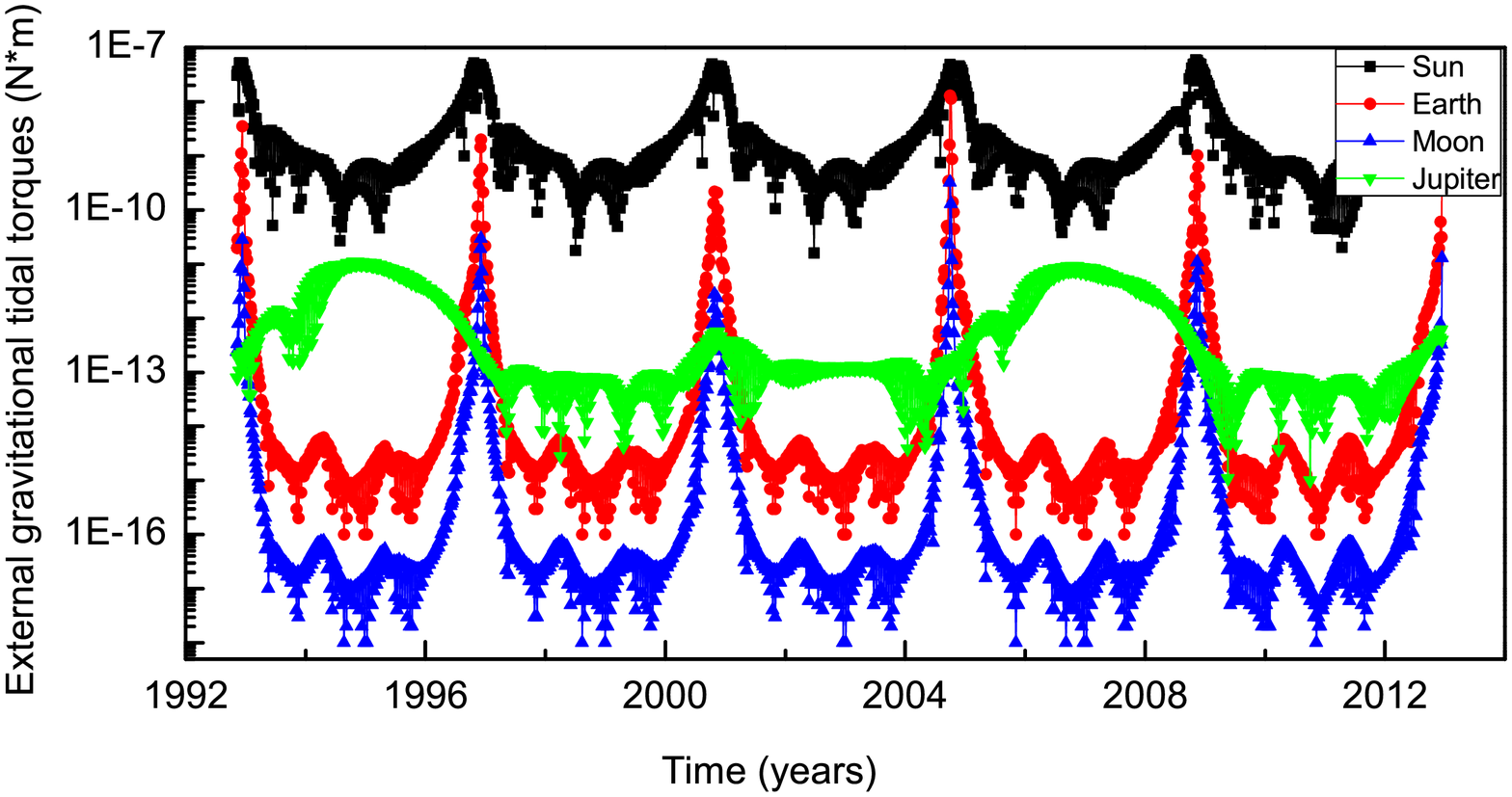}
\centering \caption{Variations in external gravitational torques.
The black, red, blue, and green curves represent the torques arising
from the Sun, the Earth, the Moon, and Jupiter, respectively. }
\end{figure*}

Based on the rotational dynamical equation and the integrated orbit,
the overall influence of these external torques on the variation in
the magnitude of Toutatis' rotational angular momentum from 1992 to
2012 was normalized with respect to the initial magnitude $H_{0}$
\citep{Takahashi2013}, as shown in Figure 10a. The terrestrial tidal
torque (see the red curve in Fig. 10b) causes a considerable change
in angular momentum when the asteroid approaches Earth at the
perihelion or during the Earth flyby that occurs every four years.
The most significant change, with a variation in angular momentum
magnitude on the order of 0.03\%, occurred in 2004 as a result of
Toutatis passing the Earth within 4.02 lunar distances. Similarly,
the tendency of the effect of the lunar torque is consistent with
that of the terrestrial torque, as shown in Figure 10c. The solar
tides always have a predominant influence on the rotational
variation. However, the terrestrial tides also play an important
role in the variation in angular momentum during Toutatis' regular
nearby visits to Earth. Figure 10d shows the influence on the
angular momentum exerted by Jupiter. The order of magnitude of this
effect slightly changed after the 2004 near-Earth flyby, and the
amplitude continually remains lower than that of the terrestrial
torque.

\begin{figure*} 
\includegraphics[scale=0.45]{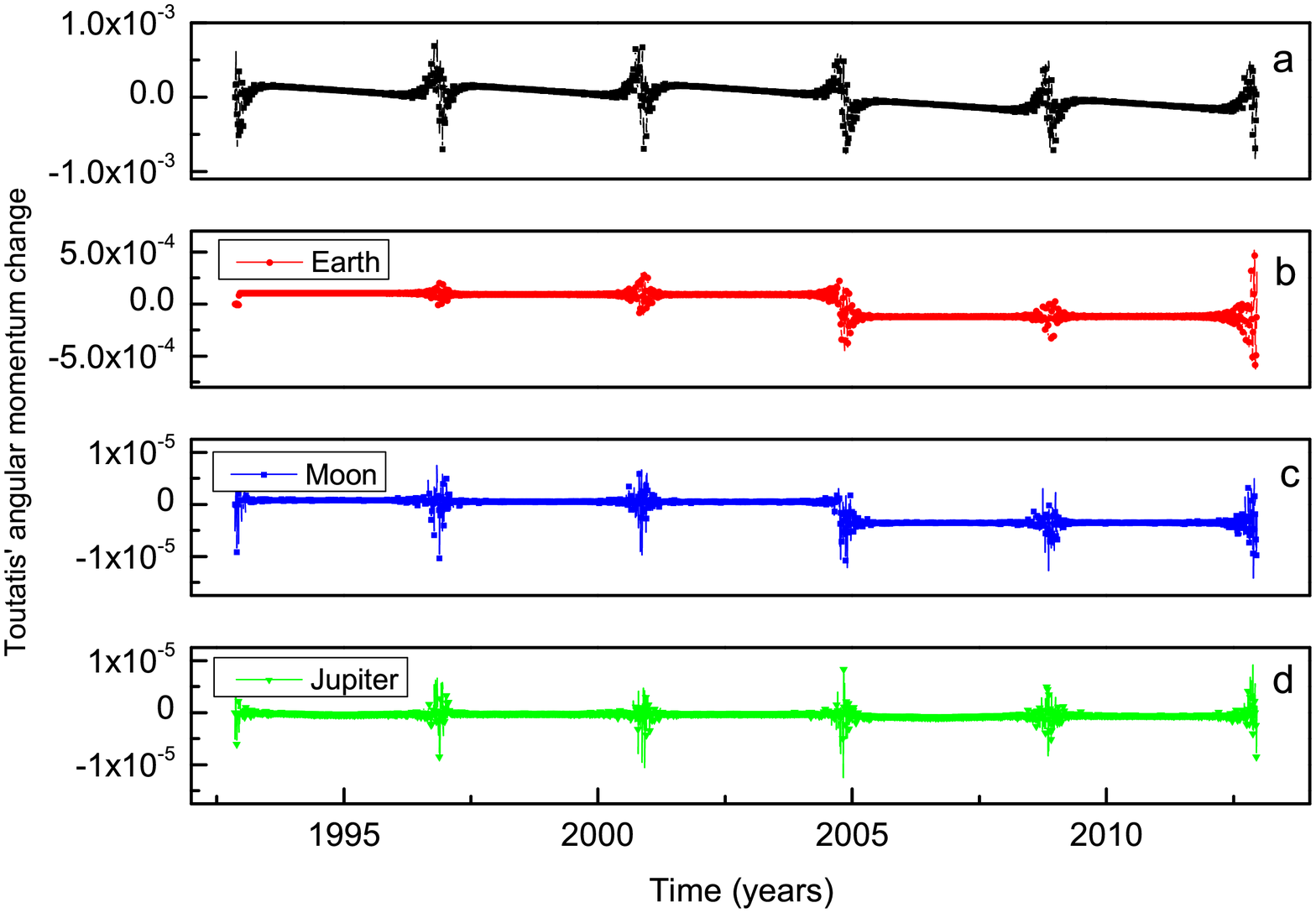}
\centering \caption{Variation in the angular momentum of Toutatis
from 1992 to 2012.}
\end{figure*}

As our simulation results indicate, the angular momentum orientation
of Toutatis is determined to be described by
($\lambda_{H}=180.2^{+0.2^\circ}_{-0.3^\circ}$ and
$\beta_{H}=-54.75^{+0.15^\circ}_{-0.10^\circ}$) and has remained
nearly unchanged in space over the past two decades. Figure 11 shows
the variations in Toutatis' angular momentum orientation from 1992
to 2012 in the J2000 ecliptic frame. The amplitude of this change is
shown to be less than one degree in both longitude and latitude.
Jumps in the angular momentum orientation occur at the perihelion of
each orbit. A small change in behavior is evident in Figure 11b as a
result of the 2004 near-Earth flyby, consistent with Figure 10b. The
variation with solar distance that is apparent in Figure 11c and 11d
indicates that the solar and terrestrial torques predominantly
affect the rotational motion of the asteroid. The misalignment of
the curves in Figure 11d is also a result of the 2004 near-Earth
flyby. Figure 11e and 11f show the angular momentum orientations of
33 sets of radar observations. Compared with the numerical results,
the observation data fall within a reasonable error range, with the
exception of a few points at large bias.

\begin{figure*} 
\includegraphics[scale=0.60]{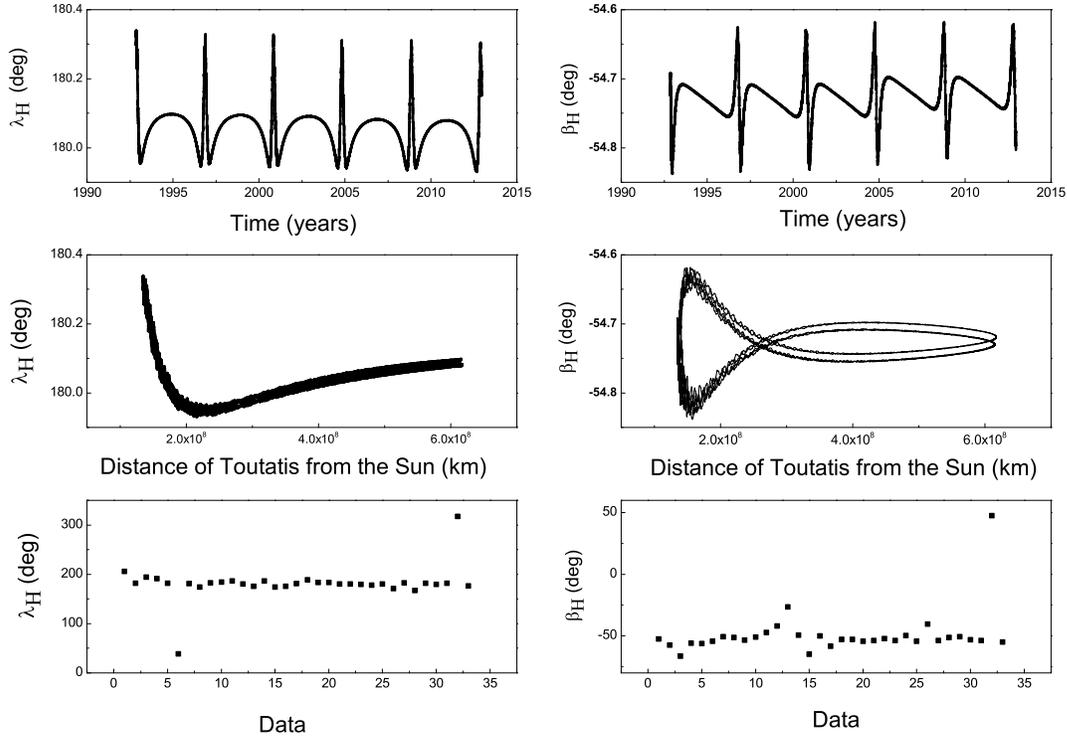}
\centering \caption{Variations in Toutatis' angular momentum
orientation from 1992 to 2012. Panels a and b show the variations in
longitude and latitude versus time. Panels c and d show the change
in longitude and latitude with the distance of Toutatis from the
Sun. Panels e and f show the corresponding results from radar
observations.}
\end{figure*}

\subsection{Rotational Period}
The motions of the short or middle axis reflect the status of the
rotation about the long axis, whereas the long axis' motion
represents precession. To calculate the two periods associated with
the spin states of Toutatis, we determined the latitudinal
variations of the asteroid's long and middle axes in the J2000
ecliptic frame, as shown in Figure 12. We applied Fourier transform
to analyze the periods of the two oscillation parameters and found
that they are 5.38 days for the rotation about the principal axis
and 7.40 days for the precession of the principal axis. These
results are in good agreement with the previous results
\citep{Ostro1999}.

Let $\beta_{X}$ and $\beta_{Z}$ indicate the latitudes of the
asteroid's long and middle axes, respectively, in the J2000 ecliptic
coordinate system. Figures 12 and 13 show the latitudinal variations
of these axes during the 1992 and 1996 flybys, respectively.
Additionally, the numerical results (represented by dotted lines in
Figs. 12 and 13) that were calculated from our dynamical model are
found to be in good agreement with the radar observations within the
error bars (marked by stars) \citep{Takahashi2013}, as listed in
Table 4. Hence, we may conclude that the orientation parameters of
Toutatis obtained from our investigation are very reliable. This
evidence provides further confirmation that our proposed rotational
model can be used to correctly evaluate the spin status of
Toutatis or other asteroids.

\begin{figure*} 
\includegraphics[scale=0.50]{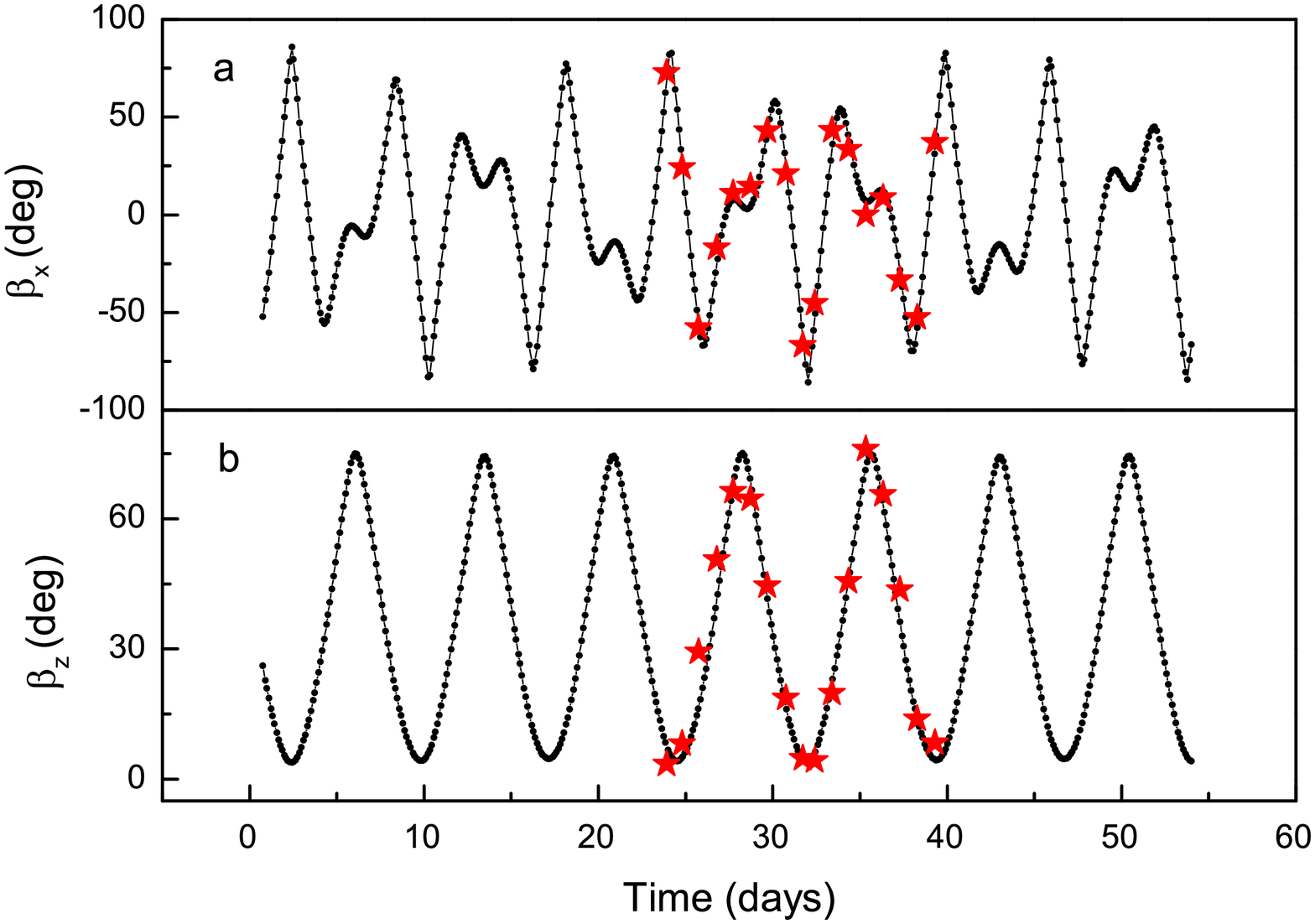}
\centering \caption{Latitudinal variations of Toutatis' long axis
$\beta_{z}$ and middle axis $\beta_{x}$ in the J2000 inertial
coordinate system in 1992.}
\end{figure*}

\begin{figure*} 
\includegraphics[scale=0.50]{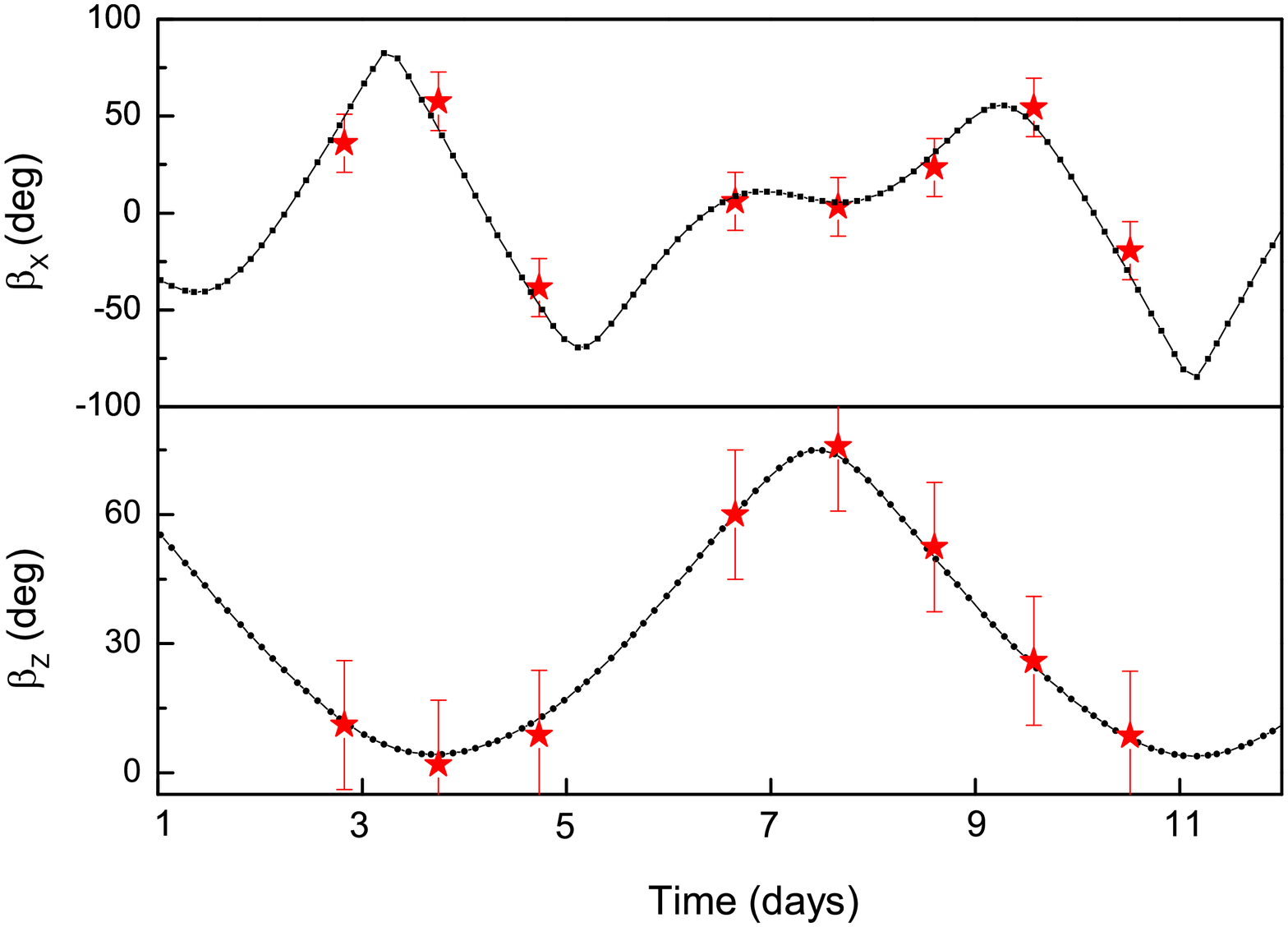}
\centering \caption{Latitudinal variations of Toutatis' long axis
$\beta_{z}$ and middle axis $\beta_{x}$ in the J2000 inertial
coordinate system during the near-Earth flyby in 1996.}
\end{figure*}

\section{Conclusions and Discussions}
In this work, we apply the observations collected during
Chang'e-2's outbound flyby to model the rotational dynamics
and determine the spin state of Toutatis. Based on
flyby images, we utilize the radar-derived shape model
to calculate Toutatis' orientation at the flyby epoch. In
addition, we estimate the 3-1-3 Euler angles to be
$-20.1^\circ\pm1^\circ$, $27.6^\circ\pm1^\circ$, and
$42.2^\circ\pm1^\circ$, respectively. Consequently, our results
have greatly improved the estimation of the orientational
parameters of Toutatis with respect to the previous predictions.

In combination with ground-based observations, we investigated the
evolution of the spin parameters using numerical simulations. In
addition to the solar and terrestrial torques, the tidal effects
arising from the Moon and Jupiter are extensively considered in our
dynamical model. The magnitude and influence of these gravitational
torques were analyzed in this work. The solar tide appears to always
be the dominant torque acting on the angular momentum of Toutatis.
Furthermore, the contribution to the external gravitation torque due
to the COM-COF offset appears to be negligible in the first-order
approximation. We also found that the closest near-Earth flyby, at
4.02 lunar distances, resulted in a 0.03\% change in the magnitude
of the angular momentum of Toutatis. The dynamical influence
exerted by Saturn on the angular momentum was also assessed in
further simulations and found to be approximately $10^2$ lower than
that of Jupiter. Hence, we can safely conclude that Saturn plays a
less important role in the variation of Toutatis' angular momentum.

The attitude at the Chang'e-2 flyby epoch that was derived from the
numerical simulations yielded a better approximation to the optical
results than that previously obtained from radar data alone. The
largest deviation in the Euler angles is observed in the pitch
angle, with a bias of less than $20^\circ$. The uncertainties
corresponding to observational uncertainties and data processing
error were considered in the simulations. The inconsistency in the
different types of observational data may have led to higher
residuals compared with previous results. Simulations based solely
on radar observations were also performed, and the corresponding rms
magnitude was much lower. However, the results obtained using a
higher-accuracy dynamical model and a combination of the various
types of observations yielded a good result that is highly
consistent with the optical images acquired during the Chang'e-2
flyby in 2012.

The precession of Toutatis was investigated by considering the
motion of its long axis. The behavior of the orbit in the inertial
frame was found to be circular, with a center axis pointing along
($-0.2^\circ, 54.6^\circ$). The precession amplitude was estimated
to be up to $60^\circ$, which may be responsible for the
significantly different attitude of the asteroid as observed by
ground-based facilities. Moreover, by exploring the motions of the
long axis and middle axis, we determined the rotation period of
Toutatis using Fourier analysis. The two major periods were found to
be 5.38 days for the principal axis rotation and 7.40 days for the
precession, in agreement with the results reported by
\citet{Ostro1999}.

Toutatis' angular momentum orientation was determined to be
described by $\lambda_{H}=180.2^{+0.2^\circ}_{-0.3^\circ}$ and
$\beta_{H}=-54.75^{+0.15^\circ}_{-0.10^\circ}$, indicating that it
has remained nearly unchanged for the last two decades. Because of
the increasing magnitudes of the solar and terrestrial torques, tiny
jumps in the angular momentum orientation occur at perihelion in
each orbital period. However, the dynamical effects caused by the
near-Earth flyby in 2004 slightly changed the latitude of Toutatis'
angular momentum orientation. Hence, our simulation results are in
good agreement with previous radar observations. In a word, based on
the combination of Chang'e-2's observations and radar data, our
investigation offers an improved understanding of the rotational
dynamics of Toutatis.

\section*{Acknowledgements}
The authors greatly acknowledge M. W. Busch and Y. Takahashi for
their helpful discussions and suggestions. This work is financially
supported by National Natural Science Foundation of China (Grants
No. 11303103, 11273068, 11473073), the Strategic Priority Research
Program-The Emergence of Cosmological Structures of the Chinese
Academy of Sciences (Grant No. XDB09000000), the innovative and
interdisciplinary program by CAS (Grant No. KJZD-EW-Z001), the
Natural Science Foundation of Jiangsu Province (Grant No.
BK20141509), and the Foundation of Minor Planets of Purple Mountain
Observatory.

\label{lastpage}

\end{document}